\documentclass[a4paper,12pt]{article}

\usepackage{graphics,amsmath,amssymb,epsfig,epstopdf}

\begin{document}

\title{
\bf Dark sector unifications: dark matter-phantom energy, dark matter - constant $w$ dark energy, dark matter-dark energy-dark matter}

\author{Dalibor Perkovi\'{c}$^{1,}$\thanks{dalibor.perkovic@zvu.hr}  and Hrvoje \v Stefan\v ci\'c$^{2,}$\thanks{hrvoje.stefancic@unicath.hr}}


\vspace{3 cm}
\date{
\centering
$^{1}$ University of Applied Health Sciences, Mlinarska street 38, 10000 Zagreb, Croatia \\
\vspace{0.2cm}
$^{2}$ Catholic University of Croatia, Ilica 242, 10000 Zagreb, Croatia }


\maketitle

\abstract{The paper brings a novel approach to unification of dark matter and dark energy in terms of a cosmic fluid.  A model is introduced in which the cosmic fluid speed of sound squared is defined as a function of  its equation of state (EoS) parameter. It is shown how logarithmic part of this function results in dynamical regimes previously not observed in cosmic fluid models. It is shown that in a particular dynamical regime the model behaves as a unification of dark matter and phantom dark energy. Further, it is shown that the model may describe dark matter - dark energy unification in which dark energy asymptotically behaves as dark energy with a constant EoS parameter larger than -1. In a specific parameter regime the unified fluid model also reproduces global expansion similar to $\Lambda$CDM model with fluid speed of sound vanishing for small scale factor values and being small, or even vanishing, for large scale factor values. Finally, it is shown how the model may be instrumental in describing the cosmic fluid dark matter-dark energy-dark matter unification. Physical constraints on model parameters yielding such transient dark energy behavior are obtained.}

\vspace{2cm}

\section{Introduction}

\label{intro}

Significant observational advances \cite{SNIa1,SNIa2,CMB1,CMB2,BAO} and ingenious theoretical attempts  in modern cosmology of the last two decades have revealed in many ways fascinating picture of the universe we live in.  The types of matter or forms of interaction known to us from terrestrial observation and experimentation seem to be able to explain only the minor part of the universe composition. Our understanding of the universe still lacks an unambiguous picture of the mechanism of the present, late-time accelerated cosmic expansion and additional matter (or modification of dynamics) responsible for the observed phenomena at the galactic and galaxy cluster levels. A most common approach is to treat these two missing elements as separate phenomena with their own specific causes. The mechanism of the accelerated expansion is predominantly contemplated within two frameworks: i) as a cosmic component, called dark energy (DE), endowed with sufficiently negative pressure in a universe in which gravity is described by general relativity (GR) \cite{DERev1,DERev2,DERev3,DERev4,DERev5} or ii) as the modified gravitational interaction without additional exotic components \cite{ModGravRev1,ModGravRev2,ModGravRev3,cadoni}. The additional matter, called dark matter (DM), has until recently been predominantly considered in the framework of weakly interacting massive particles (WIMPs). However, lack of persuasive signals for DM candidates at accelerator and underground experiments has envigorated study of alternative DM models, for some recent DM reviews and results see \cite{DM1,DM2,DM3,DM4}. Modified dynamics approaches such as MOND and its generalizations represent still actively considered alternatives to DM \cite{MOND1,MOND2,MOND3,vagnozzi}. A benchmark model describing the dark sector of the universe, thanks to its combination of simplicity and phenomenological success, is the so called $\Lambda$CDM model in which the dark energy is assumed to be the cosmological constant, whereas dark matter is assumed to be cold, i.e. nonrelativistic.  

A natural idea is to assume that dark matter and dark energy are not separate components, but two manifestations of a single unified dark component. The idea of unification of dark energy and dark matter has been implemented in numerous models, of which the most well known is the Chaplygin gas model \cite{Chap1,Chap2} and its most straightforward generalization, the Generalized Chaplygin gas \cite{GenChap1}. Since observational data on large scale structure have strongly constrained Generalized Chaplygin model \cite{GCGconstraint}, other models and extensions have been proposed. Some of these comprise Modified Chaplygins gas \cite{ModChap1,ModChap2}, Hybrid Chaplygin Gas \cite{HybridChap} and recently proposed Umami Chaplygin \cite{Umami} models. A number of approaches to unification of dark sector are based on scalar fileds: self-interacting scalar field mimicking properties of Chaplygin gas \cite{scalar1}, Dust of dark energy \cite{dustDE}, Galileon theories \cite{koutsoumbas} and k-essence \cite{kessence1,kessence2,kessence3} (for a review see \cite{scalarunified}). An interesting approach to the dark sector unification was recently proposed in the framework of superfluid dark matter \cite{superfluid}. Another dark sector unification line of study are models of interacting dark energy and dark matter. For some interesting work in this direction see e.g. \cite{Int1,Int2,Int3}. 
Furthermore, a comprehensive analysis of cosmic fluid models of dark energy was given in \cite{aether}.

In this paper we continue the study of a novel way of modeling dark energy, dark matter and their unification, first introduced in \cite{PRD2013}. In this approach the expression for the adiabatic speed of sound of the cosmic fluid $c_s^2$  (or more generally its $d p/d \rho$) is modelled as a function of its parameter of its equation of state (EoS), i.e. $w=p/\rho$. In \cite{PRD2013} a model specified by $c_s^2(w)=\alpha (-w)^{\gamma}$ was shown to lead to unified dark matter-dark energy models consistent with observations which contains Chaplygin gas and Generalized Chaplygin gas as its special cases.

This modelling approach was recently applied to the study of the cosmological constant boundary crossing and the associated phenomenon of transient dark energy \cite{Transient2018}. 
An important message from \cite{Transient2018} is that in some cases modelling in the $c_s^2(w)$ formalism should not be considered as a description of a cosmic fluid behavior, but should be used as an effective descritpion of a cosmic component in which its $d p/d \rho$ is modelled as a function of $p/\rho$.

The paper has the following structure. After the introduction presented in the first section, the second section brings the definition of the model in its explicit and implicit variant along with explicit solutions for the dynamics of $w(a)$. In the third section three interesting parameter regimes are presented: dark matter - phantom energy unification for $C=0$, dark matter - constant $w$ dark energy unification for $C=1$ and dark matter - dark energy - dark matter unification for the model with $C=1/2$. The fourth section closes the paper with conclusions.

\section{The model} 

\label{model}

In this paper we adopt the $c_s^2(w)$ modelling approach and  introduce a new model defined by
\begin{equation}
\label{eq:model}
c_s^2=w+A w \frac{ \left(-\ln \frac{w}{w_*} \right)^C}{1+w} \, ,
\end{equation}
where $A$, $w_*$ and $C$ are model parameters.
The motivation for such a choice of the model (or effectively the EoS connecting the pressure and the energy density of the component) is at least twofold.  The role of logarithmic corrections (or modifications) of the dark energy equation of state (EoS) has been recently stressed in applications of Anton-Schmidt fluid model to problems in cosmology \cite{AntonSchmidt1,AntonSchmidt2, AntonSchmidt3}. Furthermore, the model defined by (\ref{eq:model}) is analytically tractable, at least at the level of the EoS parameter as the function of the scale factor, $w(a)$. This fact allows us to reveal many physically interesting regimes and provide examples of dynamics not previously found in the literature. 

For $C \neq 0$ the model (\ref{eq:model}) can be written in a following implicit form:
\begin{equation}
\label{eq:modelimplicit}
\left[ \frac{(c_s^2-w)(1+w)}{A w}\right]^{1/C}=-\ln \frac{w}{w_{*}} \, .
\end{equation} 
This form is fully equivalent to (\ref{eq:model}) for rational $C$ if the function $x^{C}$ covers all possible branches (e.g. $x^{1/2}$ covers both $\sqrt{x}$ and $-\sqrt{x}$). In this sense, two equivalent representations  (\ref{eq:model}) and (\ref{eq:modelimplicit}) will be used interchangeably in the paper, depending which is more convenient for the discussion.

We next focus on the dynamical consequences of the model specified by the relation (\ref{eq:model}). Combining the definition for the adiabatic speed of sound $c_s^2=\frac{d p}{d \rho}$, the EoS of the barotropic fluid $p = w \rho$ and the continuity equation $d \rho + 3 \rho (1+w) \frac{d a}{a} = 0$, together with the main modelling assumption, $c_s^2=c_s^2(w)$, one obtains the dynamical equation for $w$ as the function of the scale factor $a$:

\begin{equation}
\label{eq:cs2ofw}
\frac{d w}{(c_s^2 (w)-w)(1+w)} = - 3 \frac{d a}{a} \, .
\end{equation}
Inserting (\ref{eq:model}) in (\ref{eq:cs2ofw}) and integrating yields the solution $w(a)$ for $C \neq 1$:
\begin{equation}
\label{eq:woflna}
w = w_* e^{-[3A(1-C) \ln\frac{a}{a_0}+(-ln\frac{w_0}{w_*})^{1-C}]^{\frac{1}{1-C}}} \, .
\end{equation}
Here $a_0$ and $w_0$ denote present values of the scale factor and the unified component EoS parameter, respectively. Throughout the paper we assume $w_0 < 0$. 

For $C=1$ the solution acquires the form
\begin{equation}
\label{eq:woflnaC1}
w = w_* e^{\left( \ln \frac{w_0}{w_*}\right) \left( \frac{a}{a_0}\right)^{3 A}} \, .
\end{equation}

As this model has a peculiar form, it is fascinating that it provides a very rich phenomenology. For specific values of model parameters $A$, $w_*$ and $C$, the model exhibits various forms of nonstandard dark sector unification. As the focus of this paper we single out three specific regimes:
\begin{itemize}
\item For $C=0$ the model is a realization of {\it dark matter-phantom energy} unification;
\item In the case $C=1$, the model behaves as a unified fluid exhibiting interpolation between dark matter and dark energy with constant $w=w_{*}$; 
\item The case $C=1/2$, where starting from (\ref{eq:modelimplicit}), {\it dark matter-dark energy-dark matter} unification is achieved.
\end{itemize} 

We analyze these three regimes in detail in the following section.

\section{Dynamical regimes}

\label{discussion}

The principal goal of this paper is not the exhaustive analysis of all possible regimes of the model defined by (\ref{eq:model}) (or (\ref{eq:modelimplicit})), which will be presented elsewhere, but the illustration of interesting non-standard forms of unification of the dark sector. The cases that are of particular interest in this paper are: phantom energy - dark matter unification, dark matter - dark energy unification in which the dark energy asymptotically tends to $w_{*}$ and dark matter - dark energy unification in which dark energy is a transient phenomenon in otherwise dark matter character of the unified component (which we furtheron call dark matter-dark energy-dark matter unification). We proceed with a more detailed description of these interesting regimes in the remainder of this section.  

\subsection{Phantom energy - dark matter unification}

For $C=0$ we proceed from (\ref{eq:model}). The solution (\ref{eq:woflna}) acquires an especially simple form:
\begin{equation}
\label{eq:woflnaC0}
w=w_0 \left( \frac{a}{a_0} \right)^{-3 A} \, ,
\end{equation}
whereas the expression for the speed of sound is
\begin{equation}
\label{eq:cs2C0}
c_s^2=w_0 \left(\frac{a}{a_0}\right)^{-3A} + A w_0 \frac{ \left(\frac{a}{a_0}\right)^{-3A}}{1+w_0 \left(\frac{a}{a_0}\right)^{-3A}} \, .
\end{equation}
The energy density of the unified component can be readily obtained from (\ref{eq:woflnaC0}):
\begin{equation}
\label{eq:rhoC0}
\rho= \rho_0 \left( \frac{a}{a_0}\right)^{-3} e^{\frac{w_0}{A}\left[ \left( \frac{a}{a_0}\right)^{-3 A}-1 \right]} \, ,
\end{equation}
where $\rho_0$ denotes the present value of the unified component energy density.

\begin{figure}[!t]
\centering
\includegraphics[scale=0.4]{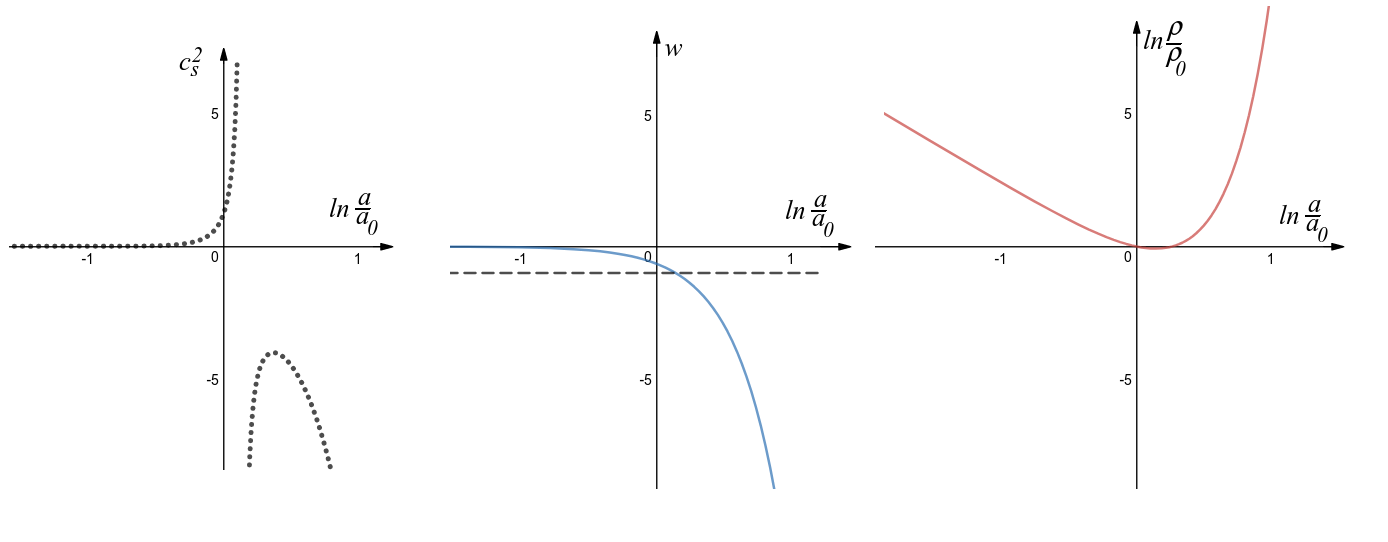}
\caption{The behavior of  speed of sound squared, parameter of EoS and energy density for $C=0$, $w_0 = -0.65$ and $A=-1$. 
}
\label{fig_C0}
\end{figure}

The expression (\ref{eq:woflnaC0}) reveals that the CC boundary crossing happens at
\begin{equation}
\label{eq:aLambda}
a_{\Lambda}=a_0 \left(-\frac{1}{w_0} \right)^{-\frac{1}{3A}} \, .
\end{equation} 
The analysis of asymptotic behavior of (\ref{eq:woflnaC0}) and (\ref{eq:cs2C0}) reveals the following:
\begin{itemize}
\item For $A > 0$: the EoS parameter has the limits $\lim_{a \rightarrow 0} w = - \infty$ and  $\lim_{a \rightarrow \infty} w = 0$. The speed of sound squared, on the other hand experiences singularity at the point of CC boundary crossing with the following limits: $\lim_{a \rightarrow 0} c_s^2=-\infty$, $\lim_{a \rightarrow a_0} c_s^2=w_0 \frac{1+w_0+A}{1+w_0}$ and  $\lim_{a \rightarrow \infty} c_s^2=0$. The limits at the singularity are $\lim_{a \rightarrow a_{\Lambda}^{-}} c_s^2= \infty$, $\lim_{a \rightarrow  a_{\Lambda}^{+}} c_s^2=-\infty$. This case is potentially interesting as a model of unification of phantom inflation at early times and dark matter at late times.

\item For $A < 0$: The limits on the EoS parameter are $\lim_{a \rightarrow 0} w = 0$ and  $\lim_{a \rightarrow \infty} w = -\infty$. The speed of sound squared is also singular at the CC boundary crossing point with the limits: $\lim_{a \rightarrow 0} c_s^2=0$, $\lim_{a \rightarrow a_0} c_s^2=w_0 \frac{1+w_0+A}{1+w_0}$ and $\lim_{a \rightarrow \infty} c_s^2=-\infty$. Finally, the limits at singuarity point are $\lim_{a \rightarrow a_{\Lambda}^{-}} c_s^2= \infty$ and $\lim_{a \rightarrow  a_{\Lambda}^{+}} c_s^2=-\infty$.

\end{itemize}
Therefore, for any sign of the parameter $A$, in one limit the dark component defined by (\ref{eq:model}) behaves as nonrelativistic matter, whereas in the opposite limit it behaves as phantom energy. From a physical point of view, the case for  $A < 0$ is especially interesting. For negative values of $A$, at small values of the scale factor (in the early universe) the dark component behaves as cold dark matter, whereas in the future it behaves as phantom energy. An illustration of such behavior is presented in Fig. \ref{fig_C0}. It is interesting to observe that for negative $A$, $c_s^2$ can be in the interval $[0,1]$ for all $a \le a_0$ if the CC barrier transition happens in the future. Therefore, interpretation of the model as a cosmic fluid can be consistent with the observational data, but singularity at CC boundary crossing and subsequent negative values for $c_s^2$ lead to instabilities in the formation of structure. Should one wish to use the model in this regime for the modelling for any $a$, it should be treated as effective modelling of $d p/d \rho$ as a function of $p/\rho$.  

\subsection{Dark matter - constant $w$ dark energy unified fluid}

Next we focus on the model behavior for $C=1$. The solution for the EoS parameter in terms of the scale factor is 
\begin{equation}
\label{eq:woflnaC1}
w = w_{*} e^{\left(\ln \frac{w_0}{w_*} \right) \left(\frac{a}{a_0} \right)^{3A}} \, ,
\end{equation}
whereas the solution for the square of the speed of sound as a function of the scale factor is
\begin{equation}
\label{eq:cs2oflnaC1}
c_s^2 = w_{*} e^{\left(\ln \frac{w_0}{w_*} \right) \left(\frac{a}{a_0} \right)^{3A}} \left[1-\frac{A \left(\ln \frac{w_0}{w_*} \right) \left(\frac{a}{a_0} \right)^{3A} }{1+w_{*} e^{\left(\ln \frac{w_0}{w_*} \right) \left(\frac{a}{a_0} \right)^{3A}}} \right] \, .
\end{equation}

The asymptotic behavior of the model is the following:

\begin{itemize}
\item For $A > 0$: The EoS parameter has the limits $\lim_{a \rightarrow 0} w = w_{*}$ and  $\lim_{a \rightarrow \infty} w = 0$, whereas the limits of the speed of sound squared are      $\lim_{a \rightarrow 0} c_s^2=w_*$ and $\lim_{a \rightarrow \infty} c_s^2=0$. 
\item For $A < 0$: The limits of the EoS parameter are $\lim_{a \rightarrow 0} w = 0$ and  $\lim_{a \rightarrow \infty} w = w_{*}$, whereas the speed of sound squared limits are 
$\lim_{a \rightarrow 0} c_s^2=0$ and $\lim_{a \rightarrow \infty} c_s^2=w_{*}$. 
\end{itemize}

\begin{figure}[!t]
\centering
\includegraphics[scale=0.4]{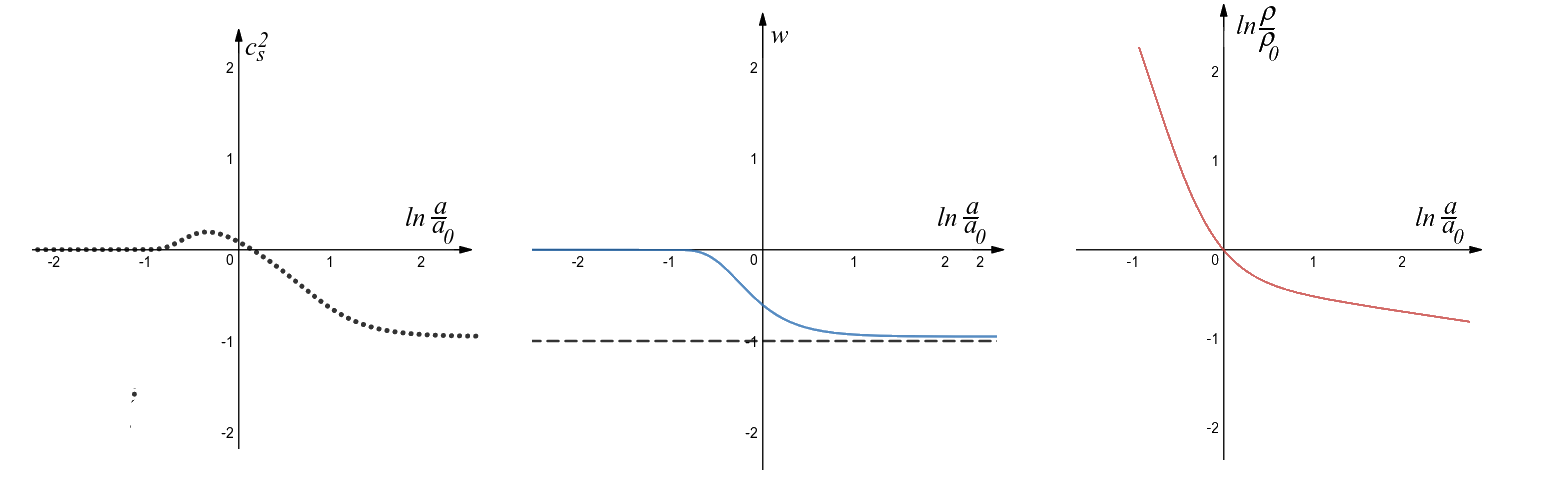}
\caption{The behavior of of speed of sound squared, parameter of EoS and energy density for $C=1$, $A=-1$, $w_0 = -0.6$ and $w_*=-0.95$. 
}
\label{fig_C1wstarabove-1}
\end{figure}

We further consider three possible situations for $w_*$: $w_*>-1$, $w_*=-1$ and $w_*<-1$. An illustrative example of model dynamics for $w_*>-1$ is presented in Fig. \ref{fig_C1wstarabove-1}. A particularly interesting feature is that for $A<0$, when the asymptotic value of the EoS parameter is a constant value $w_*>-1$. In this regime, the dynamics of $w$ never reaches $-1$, i.e. it asymptotically behaves as dark energy with a constant EoS parameter  $w_*>-1$. This result may seem contrary to claims of \cite{aether} that fluid models of dark energy behave asymptotically as cosmological constant. The authors of \cite{aether}, however, assume that $c_s^2 >0$ throughout the evolution of the cosmic fluid, whereas in our model $c_s^2$ becomes negative when $w$ approaches $w_*$. From Fig. \ref{fig_C1wstarabove-1}  it is also clear that $c_s^2$ also becomes negative when $w$ sufficiently approaches $w_*$. This fact need not be in contradiction with the observations since the transition to negative  $c_s^2$ may happen in the future and the model may describe observationally consistent unified DM-DE component. It is very intriguing to consider scenarios of future evolution of the universe after the onset of negative  $c_s^2$. In this scenario, the universe would develop strong instability to formation of gradients and the cosmic future would undergo qualitatively very different, though strictly saying nonsingular, evolution compared to the observed past dynamics.   

\begin{figure}[!t]
\centering
\includegraphics[scale=0.4]{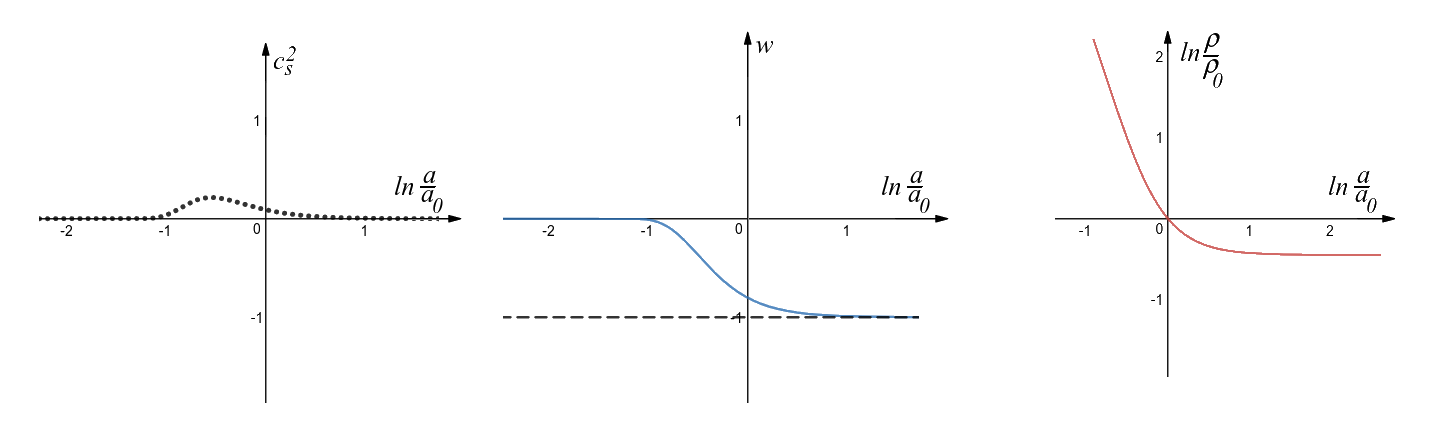}
\caption{The behavior of of speed of sound squared, parameter of EoS and energy density for $C=1$, $A=-1$, $w_0 = -0.8$ and $w_*=-1$. 
}
\label{fig_C1wstarequal-1}
\end{figure}

A physically very interesting regime, illustrated in Fig. \ref{fig_C1wstarequal-1},  emerges for  $w_*=-1$ and $A=-1$. In this case the model behaves as unified DM-DE model whichs interpolates between $w=0$ and $w=-1$. The speed of sound squared is asymptotically zero, both for distant past and distant future. Furthermore, $c_s^2$ remains between 0 and 1 for all values of the scale factor. This model is therefore potentially sufficiently similar to $\Lambda$CDM to share it phenomenological success and yet possesses distinctive features that may observationally discriminate it from $\Lambda$CDM.

\begin{figure}[!t]
\centering
\includegraphics[scale=0.3]{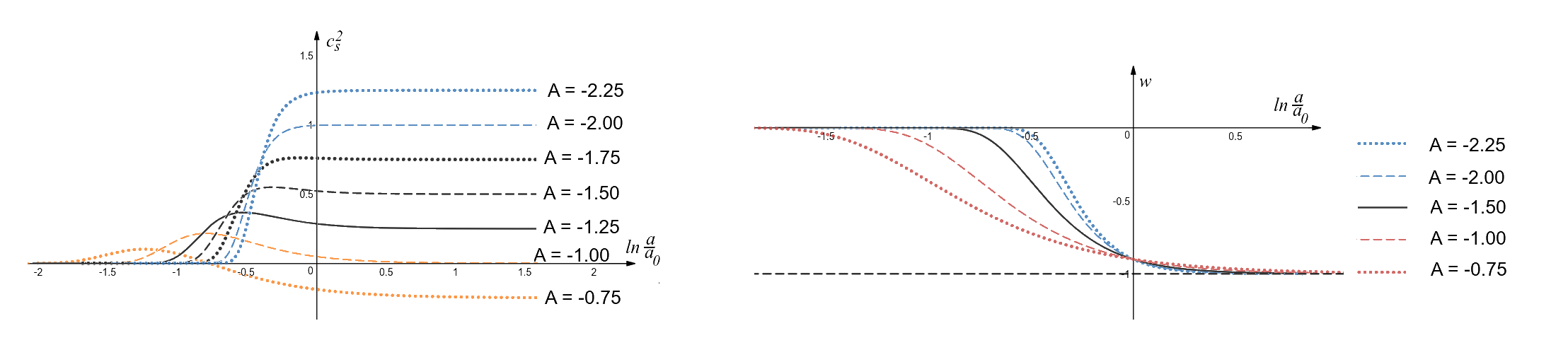}
\caption{The behavior of of speed of sound squared and parameter of EoS for $C=1$, $w_0=-0.9$ and $w_*=-1$ for various values of the parameter $A$. 
}
\label{fig_C1wstarequal-1_A}
\end{figure}

It is important to investigate if this behavior is the result of carefully chosen (or fine-tuned) parameters, or it corresponds to entire region of the model parametric space. One can see from (\ref{eq:cs2oflnaC1}) that for $w_*=-1$ it follows that $\lim_{a \rightarrow \infty} c_s^2 =-1 -A$. In Fig. \ref{fig_C1wstarequal-1_A} we present the model dynamics for $w_*=-1$ and a range of values of the parameter $A$. From the figure it is visible that $c_s^2$ remains in the interval $[0,1]$ for the entire range of the scale factor for $-2 \le A \le -1$.  In all cases in this interval of $A$, $c_s^2$ approaches $0$ for small values of the scale factor, reaches a maximum for some intermediate value of $a$ and then asymptotically reaches some positive value. For values $A>-1$, $c_s^2$ becomes negative for some values of the scale factor, whereas for $A<-2$, $c_s^2$ becomes superluminal for some $a$, as presented in Fig. \ref{fig_C1wstarequal-1_A}.

\begin{figure}[!t]
\centering
\includegraphics[scale=0.4]{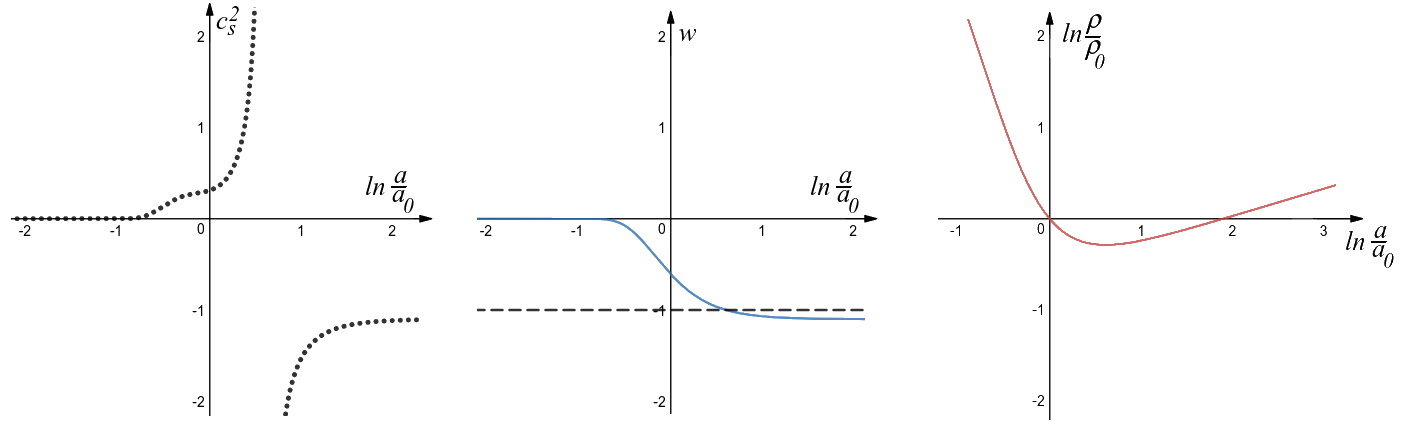}
\caption{The behavior of of speed of sound squared, parameter of EoS and energy density for $C=1$, $A=-1$, $w_0=-0.6$ and $w_*=-1.1$. 
}
\label{fig_C1wstarbelow-1}
\end{figure}

Another interesting example of dark matter - phantom energy unification can be presented for $C=1$ and $w_{*}<-1$ . Namely, from (\ref{eq:woflnaC1}) one readily obtains the following for the asymptotic  behavior:
\begin{itemize}
\item For $A > 0$: $\lim_{a \rightarrow 0} w = w_*$ and  $\lim_{a \rightarrow \infty} w = 0$;
\item For $A < 0$: $\lim_{a \rightarrow 0} w = 0$ and  $\lim_{a \rightarrow \infty} w = w_*$.
\end{itemize}
An especially interesting case is for negative $A$ where at early times the dark component behaves as cold dark matter, whereas in the asymptotic future it behaves as phantom energy with $w_{*}<-1$. An illustration of such behavior is given in Fig. \ref{fig_C1wstarbelow-1}. The problem of singularity connected to the CC boundary crossing is also present here and effective view on the model is its physically viable interpretation. 

\subsection{Dark matter - dark energy - dark matter unification}

In this subsection we focus on the model for the value $C=1/2$. In particular,  the dynamical equation for $w(a)$ starting from (\ref{eq:model}) (and assuming the branch $x^{1/2}=\sqrt{x}$)  has the form
\begin{equation}
\label{eq:dynC12}
a \frac{d w}{d a} = - 3 A w \left(-\ln \frac{w}{w_*} \right)^{1/2} \, .
\end{equation}
From the right-hand side of this expression it is easy to see that the $w(a)$ has the fixed points  at $w=0$ and $w=w_*$ and the dynamics is confined to the interval $(w_*,0)$. Consequently, the sign of $a \frac{d w}{d a}$ remains the same throughtout the range of $a$. Therefore, as the derivative of $w$ is always of the same sign, the version of the model with the explicitly defined EoS (\ref{eq:model}) cannot describe the transient dark energy solution, i.e. the sought dark matter - dark energy - dark matter unified behavior. The solution of (\ref{eq:dynC12}) is
\begin{equation}
\label{eq:woflnaC12}
w = w_* e^{-[\frac{3}{2}A \ln\frac{a}{a_0}+(-ln\frac{w_0}{w_*})^{1/2}]^{2}} \, .
\end{equation}
It is important to notice that this solution is not defined for the entire range of the scale factor $a$. In particular for $A>0$ the scale factor is in the interval $[a_*,\infty \rangle$, whereas for $A<0$ the scale factor falls in the interval $\langle 0, a_*]$, where
\begin{equation}
\label{eq:astar}
a_*=a_0 e^{-\frac{2}{3A} (-ln\frac{w_0}{w_*})^{1/2} } \, .
\end{equation}

On the other side, the dynamics for the model starting from the implicit representation (\ref{eq:modelimplicit}) is given by the following two equations
\begin{equation}
\label{eq:dynC12implicit}
a \frac{d w}{d a} = \pm 3 A w \left(-\ln \frac{w}{w_*} \right)^{1/2} \, .
\end{equation}
Whereas both dynamical equations have the same fixed points $w=0$ and $w=w_*$, the derivative of $w(a)$ in these two equations are of opposite signs. Equations (\ref{eq:dynC12implicit}) have the solutions: 
\begin{equation}
\label{eq:woflnaC12plus}
w_{+}(a) = w_* e^{- [ -\frac{3}{2}A \ln\frac{a}{a_+}+(-ln\frac{w_+}{w_*})^{1/2}]^{2}} \, ,
\end{equation}
where $w_+$ coresponds to value of $w_+(a)$ for  $a=a_+$ in (\ref{eq:woflnaC12plus}) and 
\begin{equation}
\label{eq:woflnaC12minus}
w_{-}(a) = w_* e^{- [ \frac{3}{2}A \ln\frac{a}{a_-}+(-ln\frac{w_-}{w_*})^{1/2}]^{2}} \, ,
\end{equation}
where $w_-$ corresponds to value of $w_-(a)$ for  $a=a_-$ in (\ref{eq:woflnaC12minus}). As both solutions (\ref{eq:woflnaC12plus}) and (\ref{eq:woflnaC12minus}) have $a \frac{d w}{d a}$ of the same sign as $w$ changes in the interval $[w_*,0]$, neither of them is defined in the entire interval of the scale factor $a$. In particular
\begin{itemize}
\item for $A>0$ $w_+(a)$ is defined in the interval $ \langle 0, a_*^{+}]$  and $w_-(a)$ is defined in the interval $[a_*^{-},\infty \rangle$,
\item for $A<0$ $w_+(a)$ is defined in the interval $[a_*^{+},\infty \rangle$  and $w_-(a)$ is defined in the interval $\langle 0, a_*^{-}]$,
\end{itemize}
where  
\begin{equation}
\label{eq:astarplus}
a_{*}^{+}=a_+ e^{\frac{2}{3 A} (-ln\frac{w_+}{w_*})^{1/2}} \, ,
\end{equation}
and
\begin{equation}
\label{eq:astarminus}
a_*^{-}=a_- e^{-\frac{2}{3 A} (-ln\frac{w_-}{w_*})^{1/2}} \, . 
\end{equation}
Using these two solutions, (\ref{eq:woflnaC12plus}) and (\ref{eq:woflnaC12minus}) can be written as 
\begin{equation}
\label{eq:woflnaC12plusshort}
w_{+}(a) = w_* e^{- [- \frac{3}{2}A \ln\frac{a}{a^*_+}]^{2}} \, ,
\end{equation}
and 
\begin{equation}
\label{eq:woflnaC12minusshort}
w_{-}(a) = w_* e^{- [\frac{3}{2}A \ln\frac{a}{a^*_-}]^{2}} \, ,
\end{equation}

Formally, both solutions (\ref{eq:woflnaC12plusshort}) and (\ref{eq:woflnaC12minusshort})  satisfy the equation of state (\ref{eq:modelimplicit}). Then, so does a combination of these two solutions in which the entire inerval of the scale factor is divided into nonoverlapping intervals and in each of these intervals $w(a)$ is taken to be one of solutions $w_{\pm}(a)$. In constructing such solutions one only needs to pay attention where each of $w_{\pm}(a)$ solutions is defined. The simplest version of these composite solutions for $w(a)$ is to have the full range of the scale factor divided into two subintervals with one of $w_+(a)$ and $w_-(a)$ used in one of these subintervals. Furthermore, in combining $w_+(a)$ and $w_-(a)$ we consider only continuous solutions. Depending on the choice of $a_{*}^{+}$ and $a_*^{-}$, three situations are possible:
\begin{itemize}
\item[a)] intervals where $w_+(a)$ and $w_-(a)$ are defined do not overlap;
\item[b)] intervals where $w_+(a)$ and $w_-(a)$ meet at a single point;
\item[c)]  intervals where $w_+(a)$ and $w_-(a)$ are defined do overlap.
\end{itemize}

The case in a) is not suitable for construction of model defined in the entire interval of scale factor and it will not be further considered.

\begin{figure}[!t]
\centering
\includegraphics[scale=0.3]{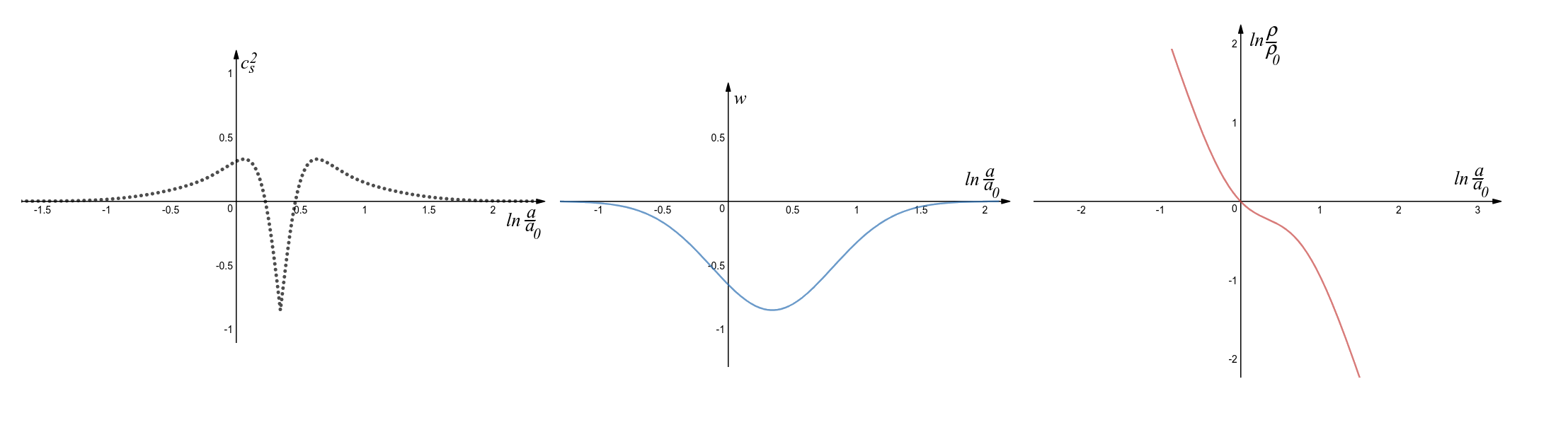}
\caption{Dark matter - dark energy - dark matter unification regime for solutions meeting in a single point. The dependence on the scale factor of speed of sound squared $c_s^2$ (left), parameter of EoS $w$ (middle) and energy density normalized by present energy density $\rho/\rho_0$ (right) for parameter values $A=-1$, $C=1/2$, $w_0=-0.65$ and $w_*=-0.85$.
}
\label{fig_DMDEDM1}
\end{figure}

The case in b) is realized for $a_{*}^{+}=a_*^{-}$ and it is depicted in Fig. \ref{fig_DMDEDM1}. It results in smooth functions for $w(a)$ and $\rho(a)$. However, in this case $c_s^2$ acquires negative values in an interval around $a_{*}^{+}=a_*^{-}$. As already mentioned in the discussion of the $C=1$ behavior, the interval with the negative values of $c_s^2$ need not be at odds with observations if it happens in future, as presented in  Fig. \ref{fig_DMDEDM1}. It is interesting to observe that if $w_{*}<-1$, the CC boundary crossing happens twice. 

\begin{figure}[!t]
\centering
\includegraphics[scale=0.25]{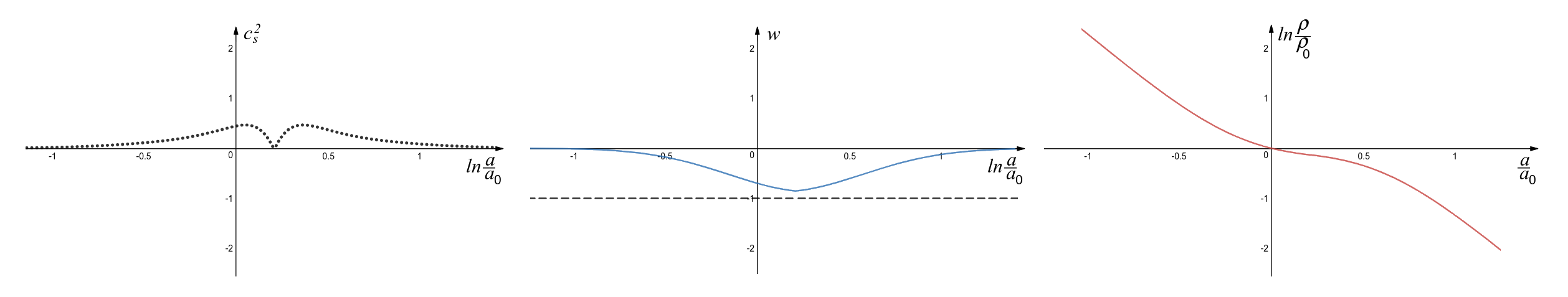}
\caption{Dark matter - dark energy - dark matter unification regime for overlapping solutions. The dependence on the scale factor of speed of sound squared $c_s^2$ (left), parameter of EoS $w$ (middle) and energy density normalized by present energy density $\rho/\rho_0$ (right) for parameter values $A=-1$, $C=1/2$, $w_{+,0}=-0.845$, $w_{-,0}=-0.72$ and $w_*=-0.88$.
}
\label{fig_DMDEDM2}
\end{figure}

\begin{figure}[!t]
\centering
\includegraphics[scale=0.5]{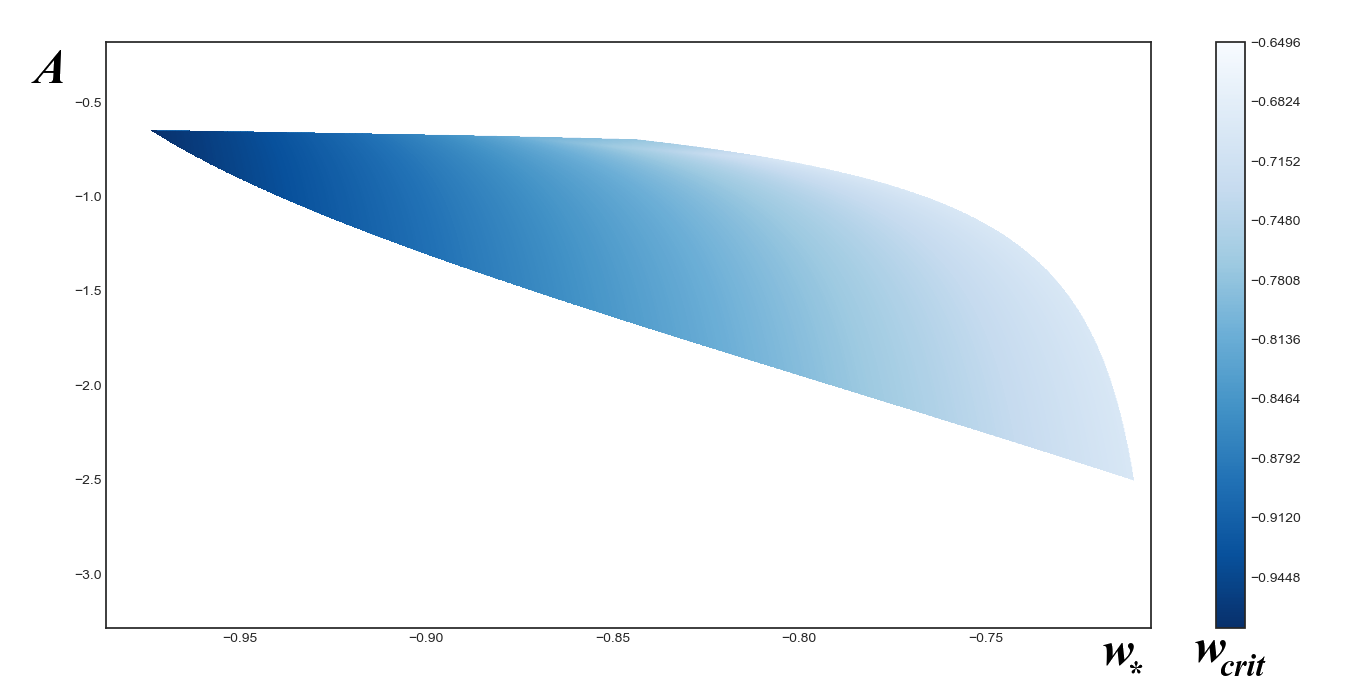}
\caption{Region of allowed parameters $w_*$ and $A$ for the DM-DE-DM dynamics with $1 \le c_s^2 \le 0$ for $w_0=-0.7$. The color scale at the right-hand side describes the minimal value of $w_{crit}$, the EoS parameter at which $w_{\pm}(a)$ meet. 
}
\label{fig_DMDEDM3}
\end{figure}

The case c)  is illustrated in Fig. \ref{fig_DMDEDM2}. The condition of continuity of the entire solution $w(a)$ requires that intervals where $w(a)$ is identified with $w_+(a)$ or $w_-(a)$ border at 
\begin{equation}
\label{eq:border}
\tilde{a} = \sqrt{a_{*}^{+} a_{*}^{-}} \, .
\end{equation}
As presented in Fig. \ref{fig_DMDEDM2}, the solution constructed in this way no longer has a smooth $w(a)$ solution, but $c_s^2$ can be made positive in the entire range of scale factor values.
An important question is if the behavior observed in Fig. \ref{fig_DMDEDM2} is the result of specifically chosen parameters, or just an illustration of a more general phenomenon. We studied this question by considering all parameters $A$ and $w_*$ for which $0 \le  c_s^2 \le 1$ for all $a$. The part of the parametric space in which this condition is satisfied for a fixed $w_0$ is presented in Fig.  \ref{fig_DMDEDM3}. In this Figure, for each combination of $A$ and $w_*$ parameters one can calculate the lowest value of $w(a)$, denoted $w_{crit}$ for which the condition is satisfied. The value $w_{crit}$ is also the value of $w$ at which two solutions $w_{\pm}(a)$ are joined. In  Fig.  \ref{fig_DMDEDM3} this value is color coded.

The solutions of the type presented in Figures \ref{fig_DMDEDM2} and \ref{fig_DMDEDM3} represent examples of transient acceleration, however without the CC boundary crossing and with $c_s^2$ in the interval $[0,1]$. Such solutions are free of future horizon problem which is present in many models in which the cosmic expansion accelerates eternally after the onset of acceleration at $z \sim 1$.

\section{Conclusions}

\label{concl}

The models of barotropic fluids defined in the $c_s^2(w)$ approach in general have an implicitly defined EoS. In this paper we have shown how a specific model defined  by (\ref{eq:model}) and its implicitly defined variant defined by (\ref{eq:modelimplicit}), in various parameter regimes, exhibit non-standard unification properties. In particular, we have demonstrated regimes of dark matter - phantom energy, dark matter - constant $w$ dark energy and dark matter- dark energy-dark matter unifications. A physically especially interesting case is realized for parameter values $C=1$, $A=-1$ and $w_*=-1$ where in the asymptotic past the model behaves like dark matter, in the asymptotic future the model behaves like cosmological constant and in the narrow intermediate region of redshift the speed of sound squared deviates from zero, still remaining in the $[0,1]$ interval.  Owing to a wealth of phenomenologically inetersting results, the model deserves a detailed comparison with the observational data which is left for future work.

Apart from concrete results for the dark sector unification as a cosmic fluid, the demonstration of three dark sector unification regimes reveals that the description of cosmic dark sector in terms of unified fluid offers even more modelling latitude than considered so far. Given that the concept of  fluid is so generic and that the true nature of neither dark matter of dark energy is established, it is both important and usefull to further exploit this latitude.

\end{document}